\documentclass[traditabstract]{aa} % for the abstract without structuration 
\usepackage{graphicx}
\usepackage{txfonts}
\usepackage{natbib}
\usepackage{url}

\bibpunct{(}{)}{;}{a}{}{,}
\newcommand{\Tx}{T_{\rm X}}
\newcommand{\Lx}{L_{\rm X}}
\newcommand{\OmM}{\Omega_{\rm M}}
\newcommand{\OmL}{\Omega_\Lambda}

\newcommand{\rhoc}{\rho_{\rm c}}
\newcommand{\Deltac}{\Delta_{\rm c}}
\newcommand{\Mdeltac}{M_{\Deltac}}
\newcommand{\xmm}{\textit{XMM-Newton}}
\newcommand{\pl}{\textit{Planck}}
\newfont{\gwpfont}{cmssq8 scaled 1000}
\newcommand{\rexcess}{{\gwpfont REXCESS}}
\newcommand{\commentaire}[1]{}

\begin{document}

   \title{The \pl\ SZ Cluster Catalog: Expected X-ray Properties}

   \author{Antoine Chamballu
          \inst{1,2}
          \and
          James G. Bartlett 
          \inst{2,3}
          \and
          Jean-Baptiste Melin\inst{4}
         }

\offprints{A. Chamballu, \email{a.chamballu@imperial.ac.uk}}

   \institute{Astrophysics group, Blackett Laboratory, Imperial College, Prince Consort road, London NW7 2AZ, UK
%              \email{a.chamballu@imperial.ac.uk}
         \and
             Laboratoire AstroParticule \& Cosmologie (APC), Universit\'{e} Paris Diderot,\\ 10, rue Alice Domon et L\'{e}onie Duquet, 75205 Paris Cedex 13, France (UMR 7164)
%             \email{bartlett@apc.univ-paris7.fr}
         \and 
             Jet Propulsion Laboratory, California Institute of Technology, 4800 Oak Grove Drive, Pasadena, CA  91109-8099, U.S.A.
         \and
         	   DSM/Irfu/SPP, CEA/Saclay, F-91191 Gif-sur-Yvette Cedex
%	   \email{jean-baptiste.melin@cea.fr}
             }

   \date{Received: July 19, 2010; accepted: May 29, 2012
   }

  \abstract{Surveys based on the Sunyaev-Zel'dovich (SZ) effect provide a fresh view of the galaxy
cluster population, one that is complementary to X-ray surveys.
To better understand the relation between these two kinds of survey, we construct an empirical cluster model 
using scaling relations constrained by current X-ray and SZ data.  
We apply our model to predict the X-ray properties of the \pl\  SZ Cluster Catalog (PCC) and compare them
to existing X-ray cluster catalogs.  We find that \pl\  should significantly extend the depth of the 
previous all-sky cluster survey, performed in the early 1990s by the ROSAT satellite, and should be particularly 
effective at finding hot, massive clusters 
($T>6\,\rm{ keV}$) out to redshift unity.  These are rare objects, and our findings suggest that \pl\  could 
increase the observational sample at $z>0.6$ by an order of magnitude.  This would open the way for 
detailed studies of massive clusters out to these higher redshifts.   Specifically, we 
find that the majority of newly-detected \pl\ clusters should have X-ray fluxes \mbox{$10^{-13}\, {\rm  ergs/s/cm}^2< 
f_X[0.5-2\; {\rm keV}]<10^{-12} \, {\rm ergs/s/cm}^2$}, i.e., distributed over the decade in flux just 
below the ROSAT All Sky Survey limit.   This is sufficiently bright for extensive X-ray follow-up 
campaigns.  Once \pl\ finds these objects, \xmm\  and \textit{Chandra} could measure temperatures to 10\% for a 
sample  of $\sim $100 clusters in the range $0.5<z<1$, a valuable increase in the number of massive
clusters studied over this range.
 }
  

   \maketitle
%
%________________________________________________________________

\section{Introduction}
The Sunyaev-Zel'dovich (SZ) effect \citep{sz:1970, sz:1972} offers a promising new
way of studying galaxy clusters, one that is complementary to the more traditional 
methods based on X-ray and optical/IR observations.
A suite of dedicated instruments is already producing high-quality SZ measurements of previously known
clusters \citep{udomprasertetal:2004, muchovejetal:2007, zwartetal:2008, basuetal:2009, halversonetal:2009, hincksetal:2009, plaggeetal:2009, 
wuetal:2009}, and the long anticipated era of SZ cluster surveys has arrived: the South Pole Telescope (SPT), Atacama Cosmology Telescope (ACT) and the \pl\ satellite are all discovering clusters solely through their SZ signal \citep{staniszewskietal:2009, vanderlinde:2010, menanteau:2010, planck:2011} 
In particular, the \pl\  satellite\footnote{http://www.esa.int/SPECIALS/Planck/index.html} is performing an all-sky SZ survey and the \pl\ consortium has recently published an early list of clusters from 10 months of observations \citep{planckVIII:2011}.  The \pl\ survey is the first 
all-sky cluster survey since the ROSAT All-Sky Survey (RASS) of the early 1990s. The ROSAT dataset produced several reference 
catalogs either directly based on the RASS, among which NORAS \citep[Northern ROSAT All-Sky galaxy cluster survey, ][]{bohringer:2000}, 
REFLEX \citep[ROSAT-ESO Flux Limited X-ray Galaxy Cluster Survey, ][]{bohringer:2004} and MACS 
\citep[Massive Cluster Survey, ][]{ebeling:2001}, or serendipitous catalogs, such as the 400sd \citep[400 Square Degree survey, ][]{burenin:2007}, 
the SHARC surveys \citep[Serendipitous High-Redshift Archival ROSAT Cluster survey, ][]{romer:2000, burke:2003} or the WARPS surveys 
\citep[Wide Angle ROSAT Pointed Survey, ][]{perlman:2002, horner:2008}. A comprehensive description of these surveys can be found in 
\citet{piffaretti:2010}. The forthcoming SZ surveys will find many new clusters in addition to numerous objects already known from X-ray (and 
optical/IR) studies.  

It is important to understand the relationship between SZ and X-ray cluster surveys for a number of reasons.  
Firstly, comparison of the two kinds of survey will clarify the nature of the selection functions of
both types of catalog.  
Secondly, comparing X-ray and SZ catalogs will yield important new scientific results on cluster physics,
including information on the SZ-mass relation crucial for cosmological applications 
of cluster evolution.  Finally, it will help dimension follow-up X-ray observations
of newly discovered SZ clusters, most notably those at high redshift.

With these objectives in mind, we develop an empirical model for X-ray and SZ
cluster signals based on observed intracluster medium (ICM) scaling relations and use it to 
predict the expected X-ray properties of SZ-detected clusters.  In the present
paper, we focus specifically on the expected \pl\  Cluster Catalog, which we refer to as the EPCC to distinguish it from the actual future. We quantify the
expected overlap between the RASS and the EPCC, and show how
the latter significantly extends the RASS cluster catalogs from $z\leq 0.3$ to redshift 
unity.  We find that newly-discovered EPCC clusters at $z>0.5$ are hot, X-ray luminous systems
with apparent X-ray fluxes falling in the decade just below the RASS sensitivity:
\mbox{$10^{-13}\, {\rm ergs/s/cm}^2< f_X[0.5-2\; {\rm keV}]<10^{-12}\, {\rm ergs/s/cm}^2$}.  

This result has important consequences for follow-up observations, implying that large numbers of \pl\  
clusters could be studied in detail with dedicated programs on \xmm\  and \textit{Chandra}.  
As an example, we show that with 25-50~ksec exposures, 
\xmm\  could  measure the temperature of EPCC clusters to 10\% out to 
$z=1$.  A 5~Msec program would then yield $\sim$~100 massive clusters with measured 
temperatures and, in the best cases, gas mass profiles in the relatively unexplored redshift range $0.5<z<1$. 
This follow-up would help calibrate the SZ-mass relation in this redshift range and
extend the reach of cluster gas mass fraction measurements and dark energy studies (e.g., \citet{galli:2012}).  
Such a project falls naturally under the category 
of  {\em Very Large Programmes} envisaged as legacy projects 
for \xmm\  after 2010.  

We begin in the next section with a presentation of our empirical cluster model and 
its observational basis.  We then discuss the EPCC, focussing on expected detection limits and 
the resulting catalog by using both analytical calculations and detailed simulations
of \pl\  observations.
Given the expected content 
of the \pl\  catalog, we apply our model to predict its X-ray properties and compare 
to the RASS.  Finally, we examine the ability of \xmm\  to follow-up in detail
a large number of newly-discovered \pl\  clusters, before concluding.

We emphasize that none of the results presented here was derived from \pl\ data. The characteristics considered here for the \pl\ survey all correspond to the pre-launch expectations, as defined in the \pl\ Blue Book \citep{planck:bluebook}. For that reason, the EPCC discussed here should be considered as representative of a \pl-like cluster catalog.

\section{Cluster Model}
\label{section:model}
We seek a simple, empirical model relating quantities that are directly observable to
each other and to cluster mass and redshift.  This is sufficient to our purpose of establishing a 
baseline model in order to:
\begin{enumerate}
\item Robustly predict the expected X-ray properties of SZ-detected clusters by direct {\em extrapolation}
	of the current observational situation;
\item Provide a reference for the interpretation of combined SZ and X-ray studies; deviations from 
	the model predictions will provide clues to missing physics, providing important feedback to more
	detailed theoretical modeling;
\item Better understand the validity of crucial scaling laws needed to use clusters as a cosmological probe, e.g., the 
          cluster counts.
\end{enumerate}
Although the model is completely general, in the present paper we focus its application on the EPCC, leaving
other SZ surveys to future work.

\subsection{Approach}

We build the model on observed scaling relations between 
X-ray or SZ observables and cluster mass and redshift, which are
the fundamental cluster descriptors.  The properties of the dark matter halos
hosting clusters are taken from numerical simulations and form the 
scaffolding on which we construct the model.  This is a standard 
approach, but we invest some time in its description to highlight its
strengths and limitations.

Clusters are essentially dark matter halos hosting hot gas, the X-ray emitting 
intracluster medium (ICM), and galaxies.  To good approximation, most 
properties are primarily functions of cluster mass and redshift, at least when
averaged over the entire cluster population.  That this is the case is perhaps not 
too surprising given that the formation of their dark matter halos
is driven by gravity alone.  The absence of a preferred scale in
gravity then implies that relations between a halo property, e.g., virial temperature, and 
mass and redshift should be power laws.  In fact, it implies that exponents of 
these power-law relations should have particular values \citep{kaiser:1986}.  
Deviations from these average
relations indicate the importance of other factors, such as accretion history 
or shape of the initial density peaks \citep{gao_white:2007, dalal:2008}.
Massive systems like clusters are rare and isolated, and present
a more homogeneous population than lower mass systems, 
such as galaxies.  This is precisely one of the reasons they are 
good cosmological probes. 

\cite{nfw:1997}, hereafter NFW, 
identified a universal dark matter density 
profile for halos from numerical simulations depending on two parameters that
can be directly related to cluster mass and redshift.  The NFW profile represents
the average dark matter distribution for halos, while of course individual objects 
scatter about this average.  We adopt the NFW mass distribution for our 
model clusters, ignoring any scatter.  This is adequate for discussing
average properties of the cluster population, but should be kept in mind
when dispersion of individual objects is under consideration.

The cluster gas is subject to physics other than just gravity, such as cooling 
and heating by feedback from member galaxies.  We could therefore expect 
the relations between gas properties and mass and redshift to differ from the 
self-similar power laws applicable to the dark matter halos.  In the more
massive clusters, however, gravity dominates the overall energetics and most of 
the gas scaling relations do not greatly deviate from their self-similar 
predictions.  Deviations tend, rather, to appear in lower mass systems ($\Tx\leq 3$~keV).
These deviations are important clues to the gas physics and hints to 
important processes driving galaxy formation.

The key ingredients of our model are:
\begin{enumerate}
\item A fiducial cosmology, which we take as the WMAP-7 best-fit model  \citep{larson:2010};
\item $\Tx - M_{500}$ relation \citep{arnaud:2005, vikhlinin:2006};
\item $\Lx - M_{500}$ relation \citep{pratt:2009};
\item $Y - M_{500}$ relation and profile for the SZ signal \citep{arnaudetal:2009};
\item A halo mass function --- we use both the \citet{jenkins:2001} and \citet{tinker:2008} mass functions. 
For the former, we need  a relation between $M_{500}$ and 
the halo mass (detailed below).   As we shall see, the two mass 
functions yield very similar predictions once normalized to the local cluster abundance.
\end{enumerate}

As discussed below in Sec.~\ref{sec:mass}, the quantity $\Mdeltac$ refers to the cluster mass within the region 
where the mean mass density is $\Deltac$ times the {\em critical density at the redshift of the cluster}, $\rho_c(z)$; the radius of this region is 
consequently referred to as $r_{\Delta_c}$. These quantities are then related by: $\Mdeltac = \frac{4\pi}{3}r^3_{\Delta_c}\rho_c(z)\Delta_c$.
The parameters of the mass function fit by Tinker et al. (2008) are given as a function of the mean overdensities $\Delta_m = \frac{\Delta_c}{\Omega_m(z)}$; as a consequence, we translate our critical overdensities $\Delta_c$ to $\Delta_m$ in order to use this mass function in the appropriate terms.

We adopt the standard flat $\Lambda$CDM WMAP cosmology with
$\OmM=0.262=1-\OmL$ and $H_0=71.4$~km/s/Mpc, taken from the WMAP seven-year analysis \citep{larson:2010}, and define 
$h_{70}\equiv H_0/(70$~km/s/Mpc$)$. We explore, however, different values for $\sigma_8$, including the WMAP-7 value 
($\sigma_8=0.801$~\citep{komatsu:2010}) as well as our best fit to the counts in the 400 Square Degree Survey (400sd) \citep{burenin:2007} with 
the observed $\Lx-M_{500}$ relation (\S\ref{section:lm}).  The latter, combined with the \citet{jenkins:2001} mass function, leads to a slightly lower $\sigma_8$, although still consistent to within $1.5\sigma$ with the WMAP-only value; the \citet{tinker:2008} mass function, on the other hand, leads to fully consistent results.

We now describe each of these ingredients in turn.

\subsection{Mass--Temperature Relation}
\label{section:mt}

Clusters are usefully described by a global X-ray temperature, although 
they are certainly not isothermal, displaying a large temperature scatter in the core 
region and a declining temperature profile in the outskirts \citep{pratt:2007}. 
X-ray observations of galaxy clusters provide both gas temperature measurements and 
total mass estimates, the latter through the assumption that the gas is in hydrostatic 
equilibrium.
\citet{arnaud:2005}, using \xmm\  observations of 10 clusters, and \citet{vikhlinin:2006}, 
using \textit{Chandra} observations of 13 clusters, independently measured the X-ray 
mass--temperature relation for {\em relaxed} systems at $z=0$.  Following these
authors, we parameterize the mass--temperature relation as
\begin{equation}
E(z)M_{500} = M_5\left(\frac{T}{5\; {\rm keV}}\right)^{\alpha_{MT}}h_{70}^{-1}
\end{equation}
where $E(z)\equiv  \frac{H(z)}{H_0} = [\OmM(1+z)^3+\Omega_{\Lambda}]^{1/2}$ and where we take for our fidicual values $M_5=4.1
\times 10^{14}M_\odot$ and $\alpha_{MT}=1.5$, consistent with both of these analyses as 
well as with self-similar scaling in cluster mass.  

\subsection{Luminosity--Mass Relation}
\label{section:lm}

\begin{table*}[t!]
\begin{center}
\caption{Luminosity--Mass relations used in the model, expressed in the form $E(z)^{-7/3}L = C\left(\frac{M}
{2\times 10^{14}h_{70}^{-1}M_{\odot}}\right)^\alpha h_{70}^{-2}$. These relations are given in~\citet{pratt:2009}, from analysis of the 
\rexcess~\citep{bohringer:2007}\, data.  All relations are corrected for Malmquist bias.  The last column gives the logrithmic dispersion of the 
luminosity relation.}
\label{tab:LMrel}
\begin{tabular}{l c c c c c}
\hline
\hline
Application & Energy band [keV] & Region [$r_{500}$] & C [$10^{44} $ erg s$^{-1}$]  & $\alpha$ & $\sigma_{\rm ln\, L,\,intrinsic}$\\
\hline
400sd distribution & [0.5-2] & 0-1.0 & $0.48\pm 0.04$ & $1.85\pm$ 0.14 & 0.414$\pm$ 0.071\\
REFLEX distribution & [0.1-2.4] & 0-1.0 & $0.78\pm 0.07$ & $1.85\pm$ 0.14 & 0.412$\pm$ 0.071\\
EPCC properties for XMM & [0.5-2] & 0.15-1.0 & $0.38\pm 0.02$ & 1.53$\pm$ 0.05 & 0.174$\pm$ 0.044\\
\hline
\end{tabular}
\end{center}
\end{table*}

This is one of the most important relations because it links the mass and redshift to the basic X-ray observable, the 
luminosity.  We will, in fact, use several $L_X-M$ relations depending on the situation, e.g., energy band considered, extent of the 
region in which the luminosity is measured, etc.  They all, however, come from \citet{pratt:2009} and were derived from the \rexcess\
sample~\citep{bohringer:2007}. 
We write the relation as follows:
\begin{equation}
E(z)^{-7/3}L = C\left(\frac{M}{2\times 10^{14} h_{70}^{-1}M_{\odot}}\right)^{\alpha_{LM}}h_{70}^{-2}
\label{eq:}
\end{equation}
with $C$ and $\alpha_{LM}$ given in Table~\ref{tab:LMrel}.
This relation shows a clear deviation from the self-similar model, where the slope $\alpha_{LM}$ 
should be 4/3. In our model, this deviation is explained by a temperature dependence of cluster
gas mass fraction consistent with a variety of observations (e.g., \citet{mohr:1999, neumann_arnaud:2001}).

\subsection{SZ Signal}
\label{sec:SZ}
The SZ signal is measured as a surface brightness change towards a cluster relative to the mean sky brightness (the cosmic
microwave background, henceforth CMB): $\delta i_\nu = y j_\nu$
where $j_\nu$ is a universal function of the observation frequency only and the amplitude is given by the Compton-y parameter 
$y(\hat{n})=\int dl \; (kT_e)/(m_ec^2) n_e \sigma_T$, an integral of the electron pressure $P = n_ekT_e$ along the line-of-sight in the direction 
given by the unit vector $\hat{n}$; the $e$ subscript refers to electron quantities and $\sigma_T$ is the Thomson cross section \citep{birk:1999}.
Because the spectral signature is universal, we can define the SZ ``flux'' in terms of the integrated Compton parameter 
$Y=\int d\Omega y(\hat{n})$, i.e., integrated over the cluster image.  This quantity has no physical units and we express it in arcmin$^2$.  

For our SZ model, we employ the gas pressure distribution $P(r)$ recently deduced from X-ray 
observations~\citep{nagaietal:2007, arnaudetal:2009} and specified as a modified NFW profile.  We use the results of \cite{arnaudetal:2009}, who 
fit a self-similar pressure profile to the \rexcess\ sample to obtain:
\begin{equation}
\label{cluster_profile}
     {P(r) \over P_{500}} = {P_0 \over x^\gamma (1+x^\alpha)^{(\beta-\gamma)/\alpha}}
\end{equation}
where $x=r/r_s$ with $r_s=r_{500}/c_{500}$ and $c_{500}=1.156$, $\alpha=1.0620$, $\beta=5.4807$, $\gamma=0.3292$.  
The normalization $P_0=8.130h^2_{70}$. The quantity $P_{500}$ expresses the scaling and is given its self-similar value:
\begin{equation}
\label{p500}
     P_{500} = 1.65 \times 10^{-3} E(z)^{8/3} \left ( {M_{500} \over 3 \times 10^{14} h_{70}^{-1} M_\odot} \right )^{2/3}  h_{70}^2  \; {\rm keV \, cm^{-3}}
\end{equation}

Assuming 0.3 solar metallicity for the ICM (the result changes very little with gas metallicity)  and integrating out to a projected distance of 
$5\times r_{500}$, we calculate the $Y-M_{500}$ relation:
\begin{equation}
\label{ym_ss_rel}
 Y \, [{\rm arcmin}^2]= Y^*_{500} \; \left ({M_{500} \over 3 \times 10^{14} h_{70}^{-1} M_\odot} \right )^{5/3} \; E(z)^{2/3} \; \left ({D_{ang}(z) \over 
 500 \, {\rm  Mpc}} \right )^{-2}
\end{equation}
where $Y^*_{500}=2.87 \times 10^{-3} {\rm arcmin}^2$ and $D_{ang}$ is the angular-diameter distance.  This pressure profile converges so that
the choice of outer radius does not greatly affect the results.   

As shown by \citet{melin:2010} and in \citet{planckX:2011, planckXI:2011}, this model provides an excellent match to direct SZ observations.  

\subsection{Redshift Evolution}

The redshift evolution of these scaling laws remains, unfortunately, quite poorly constrained, mostly as a consequence of the lack of 
high-redshift data. Recent studies are, however, consistent with self-similar  evolution of the $M_{500}-T_X$ and $L_X-T_X$ relations 
\citep{kotov:2006, pacaud:2007}. 
And in the absence of any constraints on the evolution of the $M_{500}-Y$ relation, we adopt self-similar scaling 
in this case as well.  We then have:
\begin{eqnarray}
\frac{M_{500}}{T^{\alpha_{LT}}} & \propto & E^{-1}(z)\\
\frac{L}{M^{\alpha_{LM}}} & \propto & E^{7/3}(z)\\
\frac{Y}{M_{500}^{5/3}} & \propto & E^{2/3}(z)\, D_{ang}^{-2}(z)%\, ,
\end{eqnarray}
as written explicitly in our  expressions above.

\subsection{Mass Conversion}
\label{sec:mass}

Numerous definitions of cluster mass are used in the literature, the reason being that there is no unique definition of cluster extent. 
As a consequence, our model constraints and input relations employ different mass definitions.  

We can basically divide them into three types: masses used in theoretical relations (e.g., $M_{vir}$, the mass contained within $r_{vir}$, the 
virial radius), masses used in (most of the) observed relations (e.g., $M_{500}$) and masses employed when analyzing numerical simulations 
(e.g., masses estimated by the friends-of-friends technique \citep{davis:1985}, hereafter $M_{fof}$).  The first two are defined via a chosen 
value for the density contrast $\Deltac$, defined such that $\Mdeltac$ is the mass over a region within which the mean
mass density is  $\Deltac$ times the {\em critical density at the redshift of the cluster}: $\bar{\rho} = \Deltac\, \rhoc(z)
= \Deltac \left[3H^2(z)/8\pi G\right]$.  For the virial mass, $M_{vir}$, the density contrast $\Deltac$ varies with redshift, while
in the second case it is fixed, e.g.,  at $\Deltac=500$; this is considered a good compromise between the inner regions
affected by core physics and the potentially unrelaxed regions near the virial radius 
($\Deltac\simeq180$ for an Einstein-de Sitter cosmology).

In contrast, the friends-of-friends technique establishes the perimeter of a cluster using a threshold in the distance between closest 
neighbouring particles.  This defines an isodensity contour around the cluster, without any prior on the dark matter 
distribution in the cluster itself. The isodensity contour refers to the background density in the simulation, rather than the critical
density, at the cluster redshift.

Eventually, everything needs to be expressed in terms of the same mass, say $M_{fof}$, since this is the mass used in the mass 
function \citep{jenkins:2001}.  To convert between the various mass definitions, we require a dark matter profile.  To this end, we adopt an NFW 
dark matter profile with concentration $c_{200}=4.3$ (This concentration value corresponds to the error-weighted average for the clusters used by 
\cite{vikhlinin:2006} to derive their $M_{500}-T$ relation).  Note that the conversion factor between mass definitions depends on redshift as well as 
on the underlying cosmology.  Simulations furthermore indicate that hydrostatic mass estimates are biased low due to non-thermalized bulk gas 
motions \citep{piffaretti:2008, arnaudetal:2009}.  In our mass conversion we account for a 15\% bias between the hydrostatic mass used in the 
scaling laws (e.g. $M_{500}$) and $M_{fof}$.  

\subsection{Independent Validation}
\label{section:valid}

\begin{figure}[t!]
  \resizebox{\hsize}{!}{\includegraphics{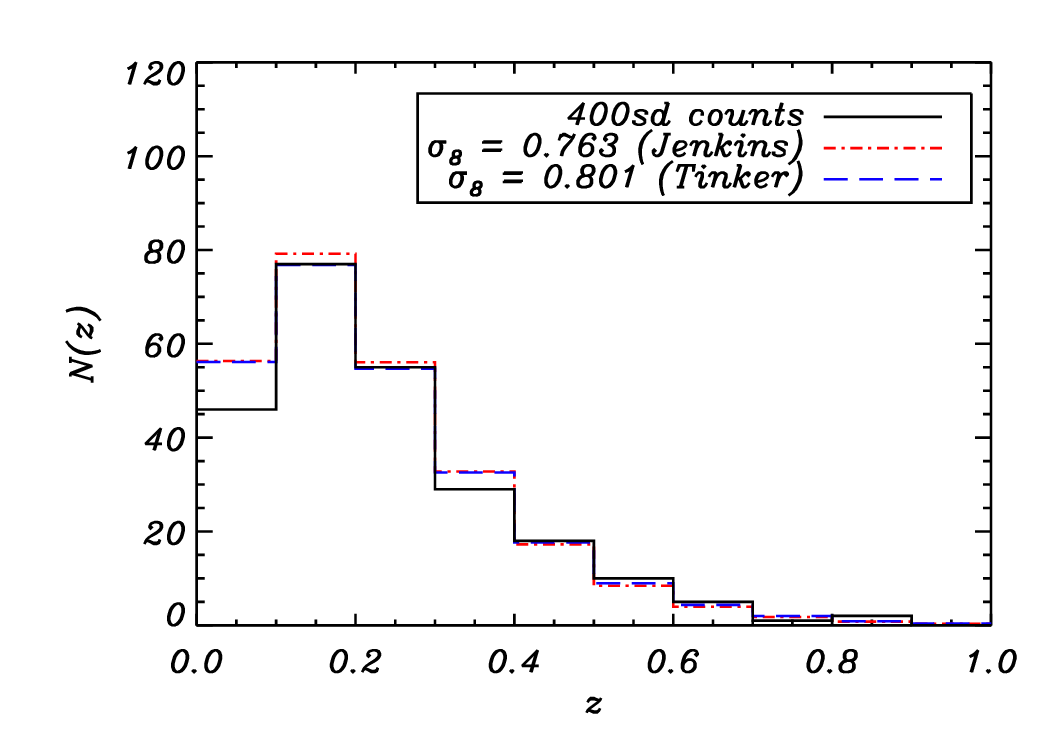}}
  \caption{Observed redshift distribution (continuous black line) in the 400sd survey~\citep{burenin:2007}, compared to the model predictions 
  for $\sigma_8=0.763$ with the \citet{jenkins:2001} mass function (dashed blue line) and $\sigma_8=0.801$ with the \citet{tinker:2008} mass 
  function (dot-dashed red line).}
  \label{fig:400deg2}
\end{figure}

\begin{figure*}[t!]
  \resizebox{\hsize}{!}{\includegraphics{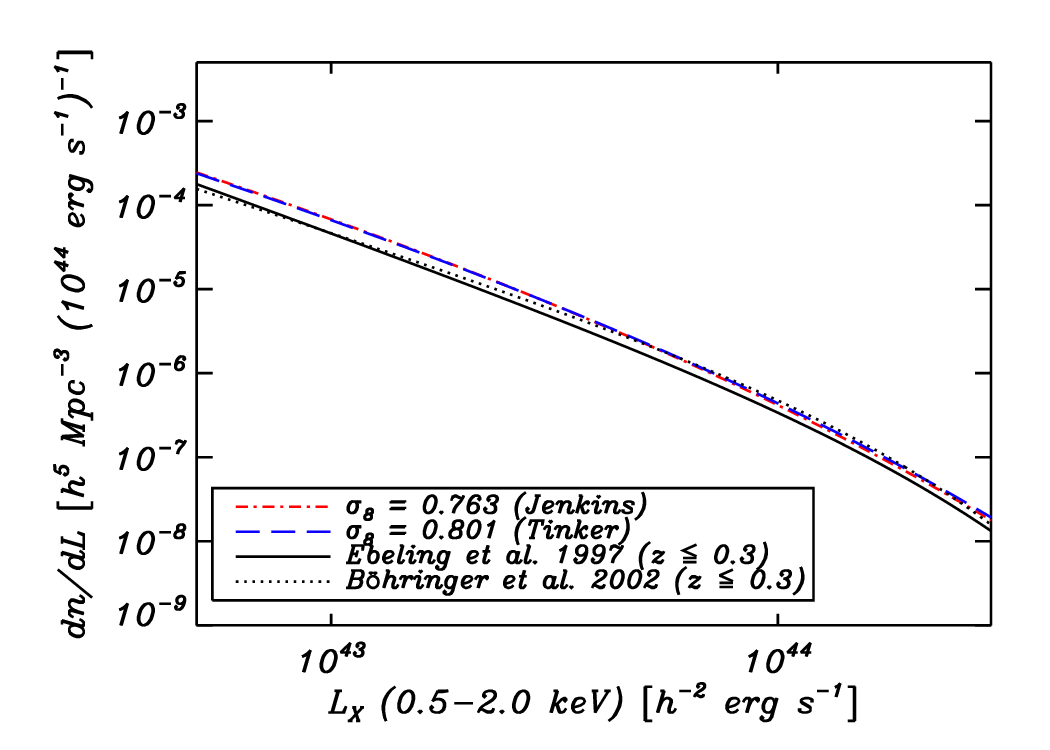}
  \hfill
  \includegraphics{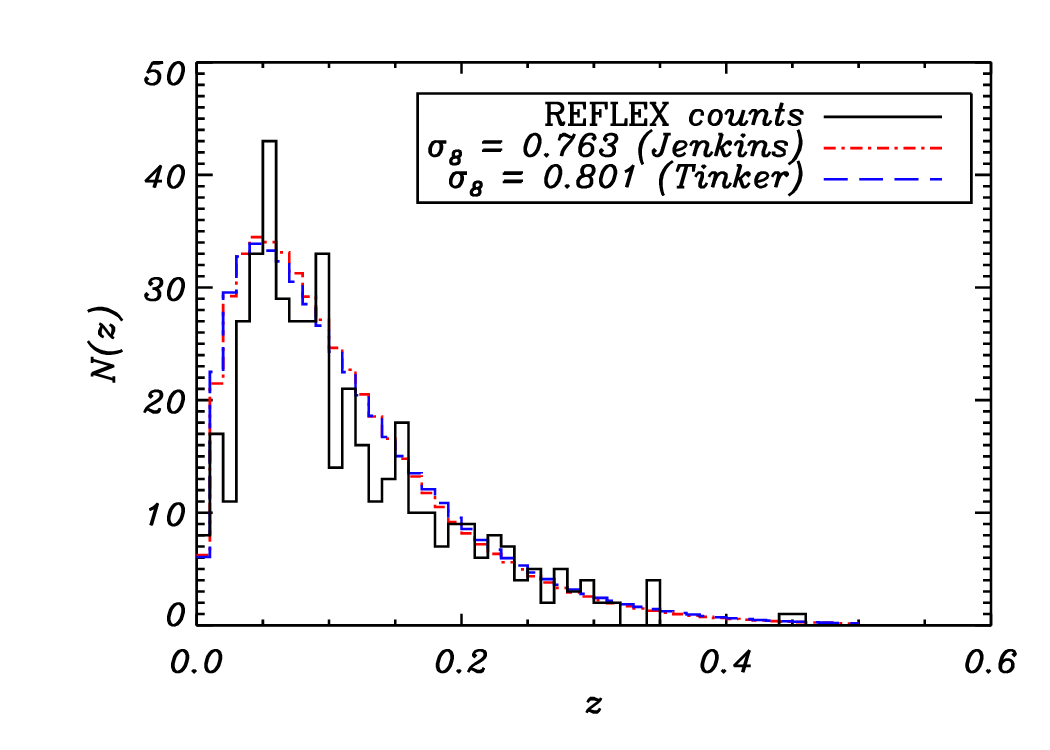}}
  \caption{Consistency checks of the cluster model. {\em Left:} Local predicted XLF for the two mass functions (along with fitted $\sigma_8$ values), compared to observations by 
  \citet{ebeling:1997} and \citet{bohringer:2002}.  {\em Right:} Same as Fig.~\ref{fig:400deg2}, but for the REFLEX survey~\citep{bohringer:2004}.
One should note that the model predictions correspond to a 100\% complete survey, while the actual survey is said to be at least 90\% complete.}
  \label{fig:reflex}
\end{figure*}

As a check of the model, we compare its predictions to other, independent observational constraints,
starting with a look at the X-ray cluster counts.  Using the $L_X-M$ relation (see Table~\ref{tab:LMrel}) and 
the \citet{jenkins:2001} mass function, we fit to the 400sd survey counts~\citep{burenin:2007} to find 
$\sigma_8 = 0.763\pm 0.008$, keeping all other paramters fixed at their fiducial values given above.  
The result is shown in Figure~\ref{fig:400deg2}.  Using the \citet{tinker:2008} mass function, we obtain $\sigma_8 = 0.801\pm 0.009$. 
Both values agree well (within 1.5$\sigma$ in the worst case) with the WMAP-7 result of $\sigma_8 = 0.801\pm 0.030$, with 
$\sigma_8 = 0.773\pm 0.025$, value derived from the SPT power spectrum~\citep{lueker:2009}, and lie within the range allowed by most recent 
measurements \citep{tyler:2004,cole:2005,reichardt:2009,juszkiewicz:2009}. Furthermore, \citet{vikhlinin:2009b} estimated the relation between 
$\sigma_8$ and $\OmM$ using clusters from the 400sd survey and the RASS. They found:
\begin{equation}
\sigma_8\left(\frac{\OmM}{0.25}\right)^{-0.47} = 0.813\pm 0.013\, ,
\label{eq:viksig8}
\end{equation}
leading to $\sigma_8 = 0.795 \pm 0.013$ when using $\OmM = 0.262$ as we did for the rest of this study, in full agreement with our estimate.
This is a non-trivial and important consistency check. 

The inclusion of the 15\% bias in mass proves indispensable, as our best estimates for $\sigma_8$, when omitting this bias, drop to 
$\sigma_8 = 0.722\pm 0.008$ and $\sigma_8 = 0.759\pm 0.008$ for the \citet{jenkins:2001} and \citet{tinker:2008} mass functions, respectively.

To follow the remaining uncertainty associated with $\sigma_8$ and the mass function in the rest of this paper, all estimates will be derived using $\sigma_8 = [0.75, 0.80, 0.85]$ for both mass functions.

Secondly, the predicted local X-ray Luminosity Function (hereafter XLF) is fully consistent with the measured XLF 
\citep{ebeling:1997, bohringer:2002}, as shown on the left-hand side of Fig.~\ref{fig:reflex}.  This is in particular true at high luminosities, i.e., for 
those clusters of greatest interest to our present study. 

Next, the model reproduces the redshift distribution of the REFLEX survey \citep{bohringer:2004}, a flux-limited survey (with 
$f_{det}=3\times 10^{-12}\, {\rm  ergs/s/cm}^2$ in the [0.1-2.4]--keV band) covering a total area of 4.24 ster in the Southern Hemisphere. The 
survey contains 447 optically confirmed clusters and claims to be at least 90\% complete.  For an equivalent survey, our model predicts a total of 
507 clusters for $\sigma_8 = 0.763$ (using the \citet{jenkins:2001} mass function), corresponding to 456 clusters when considering  90\% 
completeness and fully consistent with the observed number.  Alternatively, considering our prediction as perfect would mean that the 
completeness of the survey is 88.2\%. On the other hand, using $\sigma_8 = 0.801$ along with the \citet{tinker:2008} mass function leads to a total 
of 510 clusters, hence 459 for a 90\% completeness. A perfect prediction would, in this case, correspond to a completeness of 87.6\%. 
Furthermore, as shown in Fig.~\ref{fig:reflex}, the overall shapes of the simulated and observed redshift distributions are in both cases in very good 
agreement; note, however, that the completeness is not taken into account in this plot since the redshift of the missing clusters is not known. 

Finally, concerning the SZ part of our model, we have already noted that the model provides an extremely good accounting of direct SZ observations
\citep{melin:2010, planckX:2011, planckXI:2011}.  We hence consider the model to be 
a reasonable empirical representation of the current data and will now apply it to study the expected X-ray properties of SZ-detected clusters.  We 
emphasize that the model can be very easily adapted to new observational and theoretical contexts: it is built around few analytical ingredients 
that can be trivially changed if needed, which is an important strength of this approach.

\section{SZ surveying and the \pl\ cluster catalog}

Like the X-ray emission, the SZ effect identifies clusters through the 
presence of the hot, ionized intracluster medium (ICM).  The SZ effect, however, is 
less sensitive to gas substructure, being directly proportional to the thermal energy 
of the ICM, and remains unaffected by cosmic dimming.  As a consequence, 
SZ cluster surveys efficiently find clusters at high redshift. 

The \pl\ satellite is the third generation space mission dedicated to 
the CMB, following COBE and WMAP.  With its higher sensitivity, angular
resolution and wide frequency coverage, it is the first capable of finding
large numbers of galaxy clusters through the SZ effect.  Its all-sky catalog 
of clusters detected through the SZ effect (the \pl\ Cluster Catalog, or PCC)
will be one of the primary and novel scientific products of the mission.
An early version of the catalog was recently published in \citet[ESZ]{planckVIII:2011} based on 
high signal-to-noise detections in the first tens months of data.  
The \pl\ catalogs (ESZ and Legacy catalogs) are, remarkably, the first all-sky cluster catalog since those
resulting from the ROSAT All-Sky Survey (RASS) of the early 1990s~\citep{trumper:1992}.
They are expected to become a workhorse for many cluster and cosmological
studies.

\subsection{Properties of the Expected \pl\ Cluster Catalog (EPCC)}

We use the matched-multifilter~\citep{herranz:2002} detection algorithm described by~\citet{melin:2006} to constitute 
the EPCC.  Taking a set of sky maps at different frequencies, the algorithm simultaneously 
filters in both frequency and angular space to optimially extract objects with the thermal SZ 
spectral signature and the expected SZ angular profile.  The filter is applied with a range of 
angular scales to produce a set of candidates at each scale, which are subsequently merged 
into a single catalog.  Construction of the filter requires adoption of a spatial 
template, for which we use the modified NFW pressure profile described in Eq.~(\ref{cluster_profile}).  In essence,
we assume that the filter template exactly matches the true cluster profile, which should be
borne in mind as an idealized situation. 

\begin{figure}[t!]
\hspace{0.7cm}
 \resizebox{7.3cm}{7.3cm}{\includegraphics{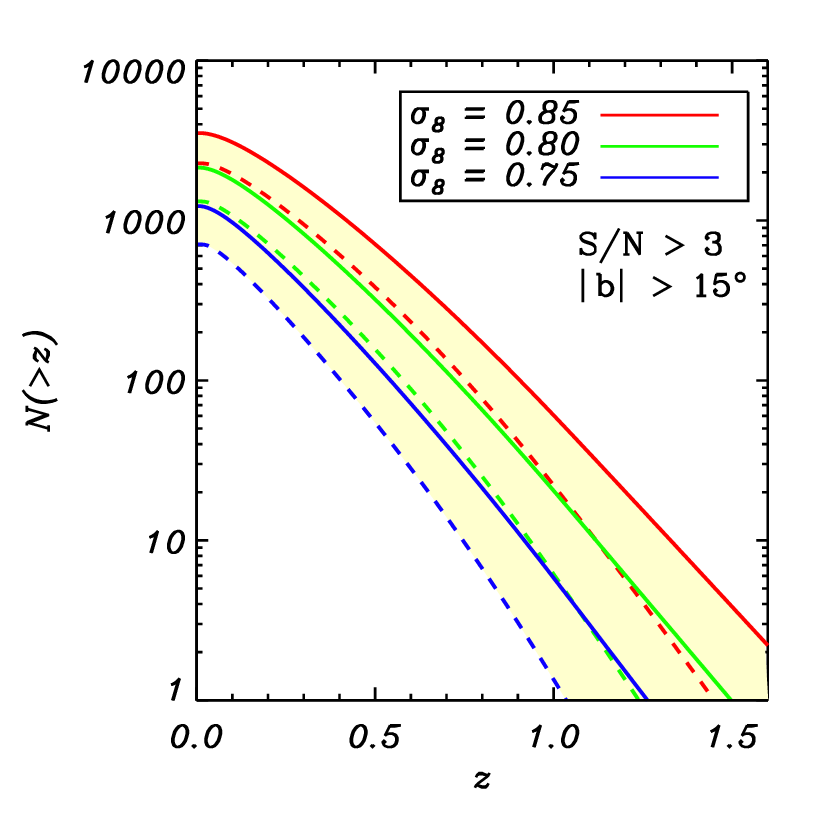}}
  \caption{Predicted cumulative redshift distribution of the EPCC for a nominal mission length 
  (14 months) and threshold $S/N>3$.  The beige band illustrates modeling uncertainty due 
  to $\sigma_8$ and the choice of the mass function. The solid lines are derived using the \citet{jenkins:2001} mass function while the dashed lines 
  correspond to the \citet{tinker:2008} mass function.}
  \label{fig:pcc}
\end{figure}

We apply our filter to simulations of the \pl\ data set based on an early version of the \pl\
Sky Model (PSM\footnote{{http://www.apc.univ-paris7.fr/~delabrou/PSM/psm.html}}).  These simulations 
provide a set of frequency maps including primary CMB temperature anisotropies in a WMAP-5 
only cosmology; Galactic synchrotron, free-free, thermal and spinning dust emission; 
extragalactic point sources; scan-modulated instrument noise and beams \citep{planck:bluebook}.  
Note, however, that we do {\bf not} include any clusters in the simulations: by 
running our filter over these simulated maps, we seek only to quantify the total filter noise 
due to all instrumental and astrophysical sources.  

This gives us our noise threshold at each filter scale as a function of position on the sky.  
With this information and our SZ scaling law, we then calculate the detectable 
cluster mass at a given signal-to-noise threshold as a function of redshift, over the sky 
and for each filter scale. By using the \citet{jenkins:2001} and \citet{tinker:2008} mass functions in our adopted
cosmology, we find the EPCC distribution in mass and redshift.
Figure~\ref{fig:pcc} shows the resulting cumulative redshift distribution for $S/N>3$ at Galactic latitudes
$|b|>15^{\circ}$ for a nominal 14-month mission, comprising two full-sky surveys, for three $\sigma_8$ values that illustrate the uncertainties on 
the determination of its value. It should be noted though, as shown above, that the cases that correspond best to the local counts are $\sigma_8 = 0.75$ for the \citet{jenkins:2001} and $\sigma_8 = 0.80$ for the \citet{tinker:2008}.
Our results here are consistent with the more extensive study of a dozen detection algorithms run on 
simulated \pl\ data and presented in Melin et al. (2012, in preparation).

\subsection{Comparison with the RASS}

\begin{figure}[t!]
  \resizebox{\hsize}{!}{\includegraphics{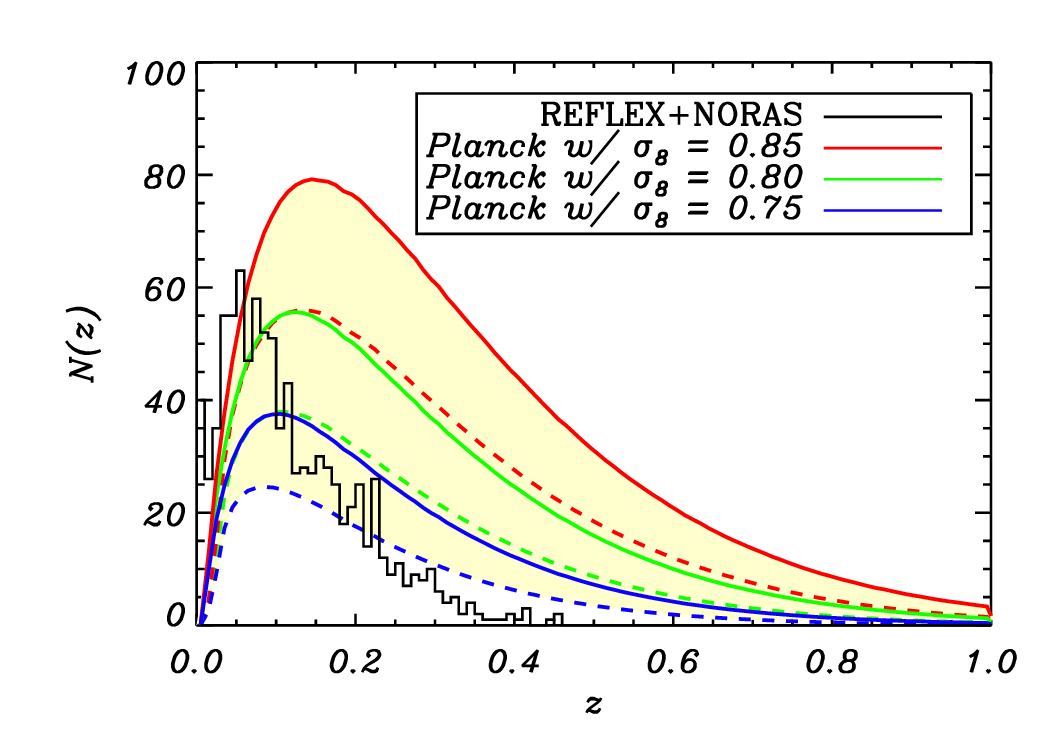}}
  \caption{Comparison of the ROSAT cluster catalog (REFLEX+NORAS; black) and EPCC differential redshift distributions, shown 
  for different values of $\sigma_8$. The solid lines are derived using the \citet{jenkins:2001} mass function while the dashed lines correspond to 
  the \citet{tinker:2008} mass function. The EPCC is expected to significantly extend the depth of all-sky cluster surveys.}
  \label{fig:pl-ros}
\end{figure}

Since the \pl\ survey will be the first all-sky cluster survey since the RASS, it is of great interest to compare the EPCC to 
the ``ROSAT Cluster Catalog'', by which we mean the sum of the REFLEX and NORAS catalogs~\citep{bohringer:2000,bohringer:2004}.  
This is done in Fig. \ref{fig:pl-ros}, which 
shows the measured ROSAT Cluster Catalog redshift distribution and the EPCC distribution for our three values 
of $\sigma_8$.  From this comparison we see, as could be expected, that many clusters are common to both surveys.  
One of the first products of the
\pl\ Cluster Survey will therefore be the detailed study of the relation between SZ and X-ray properties afforded by the joint
catalog.  Such a study will furthermore benefit from the fact that the RASS clusters are generally well-studied, allowing a number of 
important scaling relations to be derived.  This will lead, for example, to additional tests of our empirical model and eventual modifications, 
all providing useful insight into cluster physics.  

Moreover, prior knowledge of the X-ray-derived characteristics of these joint clusters will enable more precise recovery of their SZ properties. This 
is notably the case for cluster angular size, which is difficult to recover from the \pl\ data alone due to the size of the effecive SZ beam.  This 
difficulty translates into noticably larger SZ flux ($Y_{SZ}$) uncertainties than if cluster size were known a priori~\citep{melin:2006} (also, Melin et 
al. 2010, in preparation). Clusters in the joint \pl-ROSAT catalog will therefore have significantly more precise SZ flux measurements thanks to the 
more precise X-ray information on cluster extent.

More surprisingly, perhaps, is the fact that some RASS clusters {\em are not}  seen in the EPCC.  These tend to be mostly local and extended 
clusters that are ``resolved-out'' by the \pl\ survey.  As noted in \citet{melin:2005, melin:2006}, the \pl\ selection function depends on both cluster 
flux and extent.  The fact that the \pl\ selection curve cuts through the observed RASS clusters is extremely important: it will be invaluable in 
constraining the PCC selection function by providing well-studied clusters that fall outside of the PCC and thereby improve our 
understanding of the selection criteria.  This is a fortunate and unusual situation that we will fully exploit when constructing the PCC. 

Finally, as can be seen, we expect many new clusters not seen in the RASS.  The PCC should greatly extend the redshift reach of all-sky
cluster surveys over that of the RASS.  The next section is dedicated to these new \pl\ clusters.

\subsection{\xmm\ follow-up of the new \pl\ clusters}

\begin{table*}[ht!]
\begin{center}
\caption{Comparison between the predicted numbers of clusters in different subsets in the EPCC for different $\sigma_8$ values.}
\label{tab:numbers}
\begin{tabular}{p{7cm} c c c | c c c}
\hline
\hline
Characteristics & \multicolumn{6}{c }{Predicted number of clusters}\\
\cline{2-7}
 & \multicolumn{3}{c }{Jenkins} & \multicolumn{3}{c }{Tinker}\\
\cline{2-7}
& $\sigma_8 = 0.75$ & $\sigma_8 = 0.80$ & $\sigma_8 = 0.85$ & $\sigma_8 = 0.75$ & $\sigma_8 = 0.80$ & $\sigma_8 = 0.85$\\
\hline
All & 1\,229 & 2\,140 & 3\,523 & 709 & 1\,316 & 2\,277 \\
New clusters & 425 & 853 & 1576 & 220 & 480 & 945 \\
$0.8\geqslant z>1.0$ & 15 & 44 & 110 & 5 & 18 & 53\\
$z\geqslant 1.0$ & 6 & 20 & 60 & 1 & 6 & 21\\
$T\geqslant 6\rm{\ keV}$; $z\geqslant 0.5$ & 122 & 306 & 674 & 50 & 146 & 357 \\
$T\geqslant 6\rm{\ keV}$; $0.8\geqslant z>1.0$ & 15 & 44 & 110 & 5 & 18 & 53\\
$T\geqslant 6\rm{\ keV}$; $z\geqslant 1.0$ & 6 & 20 & 60 & 1 & 6 & 21\\
$T\geqslant 6\rm{\ keV}$; $0.8\geqslant z>1.0$; $f_x > 10^{-13}\, {\rm  ergs/s/cm}^2$ & 15 & 44 & 110 & 5 & 18 & 53\\
$T\geqslant 6\rm{\ keV}$; $z\geqslant 1.0$; $f_x > 10^{-13}\, {\rm  ergs/s/cm}^2$ & 5 & 19 & 55 & 1 & 5 & 20\\
\hline
\end{tabular}
\end{center}
\end{table*}

\begin{figure*}[t!]
  \resizebox{\hsize}{!}{\includegraphics{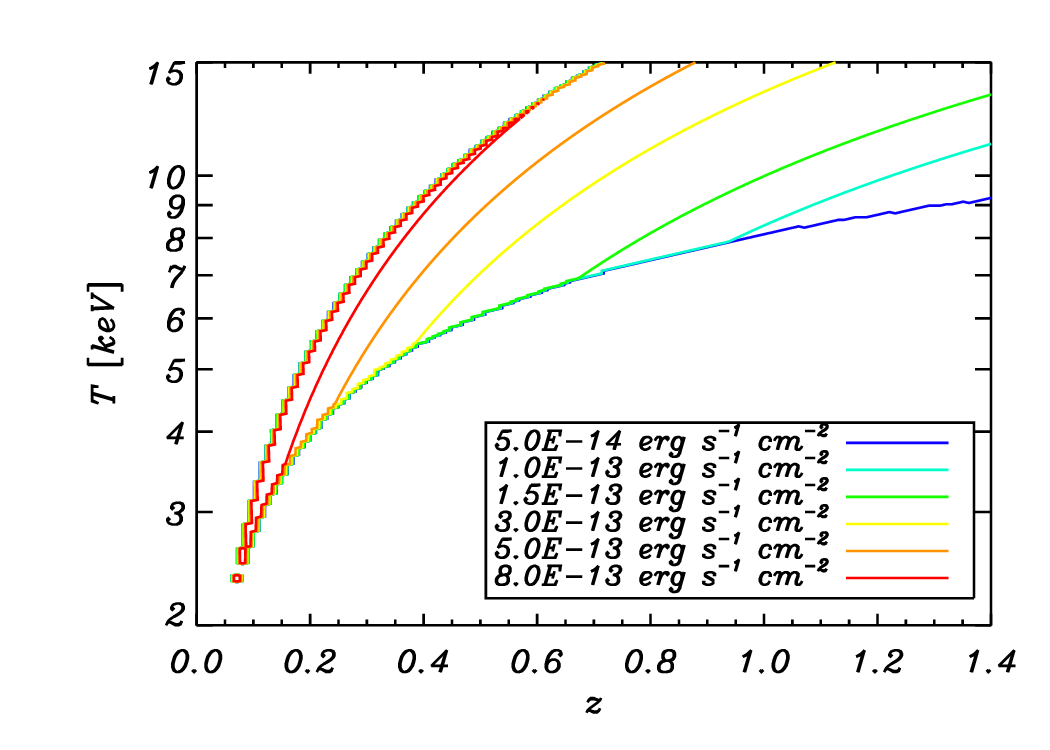}
\hfill
  \includegraphics{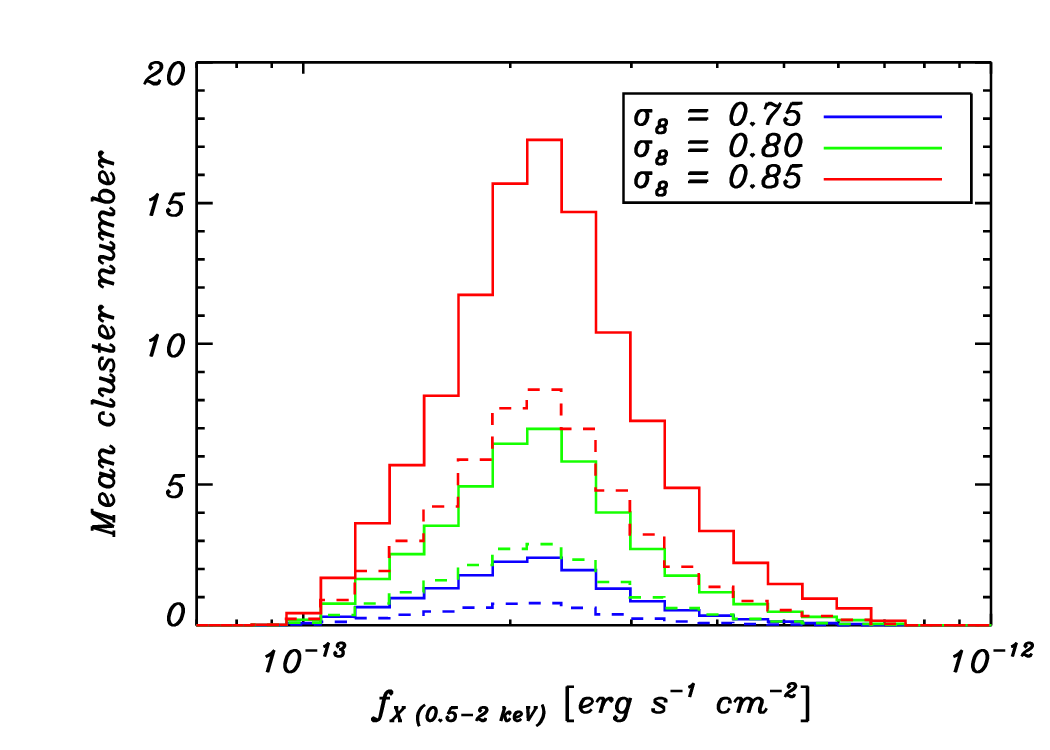}}\\
  \caption{X-ray characteristics of new EPCC clusters within $[0.15-1.0]\times r_{500}$.  {\em Left:} Contours of iso-flux in the \xmm\ [0.5-2]--keV band  
  plotted in the ($z, T$) plane.  {\em Right:} X-ray flux distribution of new high-redshift ($z\in [0.8-1]$) clusters for $\sigma_8=0.75$ (blue), 
  $\sigma_8=0.80$ (green) and 
  $\sigma_8=0.85$ (red). The solid lines correspond to results obtained using the \citet{jenkins:2001} mass function while the dashed line was 
  obtained using the \citet{tinker:2008} mass function.}
  \label{fig:contours}
\end{figure*}

We define \textit{new} \pl\ clusters as those with a predicted X-ray flux in the ROSAT [0.1,2.4]--keV band less than $f_{\rm X}=3\times 10^{-12}\, {\rm  
ergs/s/cm}^2$, which corresponds to the REFLEX detection limit and the deepest detection limit of the NORAS survey; we consider that all the 
other clusters are contained in the RASS catalogs.  We then use the model to predict the X-ray properties of these new clusters in the \xmm\ 
[0.5,2]--keV band. Fig. \ref{fig:contours} shows the contours of isoflux within $[0.15-1.0]\times r_{500}$ in the \xmm\ band over the ($z, T$) plane for 
these new clusters.

The key point is that we expect a substantial fraction of these new clusters to be hot (or equivalently, massive) and distant; in other words, those
clusters that are the most useful cosmological probes.  Table \ref{tab:numbers} lists quantitative details on the distribution of these clusters. It 
particularly focusses on hot (i.e. with $T>6$ keV) and distant (i.e. with $z>0.5$) clusters as they fall in the most poorly filled region of the $(T,z)$ 
plane for known clusters.
Moreover, these clusters are predicted to be very luminous, with X-ray flux in the \xmm\ band greater than $10^{-13}\, {\rm ergs/s/cm}^2$ out to 
redshifts of order unity. 

The following expression provides a useful estimate of the exposure time  $t_{exp}$ (in ks) needed to measure a global temperature to 
$\sim 10 \%$ given its X-ray flux in the XMM band\footnote{M. Arnaud, private communication}:
\begin{equation}
t_{exp} = 55\times\left(\frac{f_x}{10^{-13}{\rm\ ergs/s/cm}^2}\right)^{-1.35}\, ,
\label{eq:texp}
\end{equation}
We see that an exposure time of 55~ks is sufficient to obtain a temperature measurement for any cluster in the 
EPCC.  Fig. \ref{fig:contours} shows, as illustration, the histrogram of the number of new clusters with $0.8\geqslant z>1.0$  
in the EPCC as a function of flux in the \xmm\ band within $[0.15-1.0]\times r_{500}$.  All of the clusters have fluxes lying in the decade below the 
RASS detection limit and are thus easily observable with \xmm\ or \textit{Chandra}.  

In order to get an idea of the capabilities of \xmm\ to observe such clusters based on actual measurements, we refer to the observations made 
of two of the four known clusters with $T\geqslant 6\rm{\ keV}$ and $0.8\geqslant z>1.0$ and consider the results obtained and required 
exposure times:

\paragraph{MS 1054-0321 \citep{gioia:2004}:} The \xmm\ observations yield temperature, redshift (through observation of the iron 
line in the X-ray spectrum) and flux estimates of $T = 7.2^{+0.7}_{-0.6}$ keV, $z = 0.847^{+0.057}_{-0.040}$ and 
$f_{{\rm X\ }[0.5,2]{\rm\ keV}} = (1.9\pm 0.09)\times 10^{-13}{\rm\ ergs/s/cm}^2$, respectively; and they also provide 
a detailed image showing several components to the cluster's structure.  The effective exposure time needed was $\sim$25 ks.
\paragraph{RX J1226.9+3332 \citep{maughan:2007}:} The same quantities were measured in this case: $T = 10.4\pm 0.6$ keV, 
$z = 0.89$ and $f_{{\rm X\ }[0.5,2]{\rm\ keV}} = 3.27\times10^{-13}{\rm\ ergs/s/cm}^2$ \citep{vikhlinin:2009a}.  Moreover, the observation of the 
temperature profile, combined with the assumption of hydrostatic equilibrium, enabled a determination of the total mass and gas mass profiles, 
and thus an estimate of the cluster mass: $M_{500} = 5.2^{+1.0}_{-0.8}\times 10^{14}M_{\odot}$. In this case, the exposure time was $\sim$70 ks.\\

These clusters are good examples of the kind of objects we find in the EPCC.  We conclude therefore that 25-50 ks exposures 
is sufficient to obtain temperature measurements to $\sim 10\%$ for any cluster in the EPCC, even out to $z\sim 1$.  It is even
sufficient, in some cases, to obtain mass estimates.  Detailed X-ray studies of a large fraction of the new \pl\ clusters is hence feasible
and would significantly advance our understanding of cluster structure at intermediate to high redshifts.

\section{Conclusions}
The \pl\ SZ survey will be the first all-sky cluster survey since the workhorse RASS dating from the early 1990s~\citep{trumper:1992}.  
The \pl\ Cluster Catalog (PCC) will
thus satisfy the long-standing need for a deeper all-sky cluster survey.  We have constructed an empirical model for cluster SZ and X-ray 
properties that incorporates the latest results from detailed X-ray and SZ studies.  Based on this model, we expect the PCC to extend the RASS 
to redshifts of order unity, as shown in Figs.~\ref{fig:pcc} and \ref{fig:pl-ros}, and to contain a high fraction of hot, massive and luminous clusters.  
These objects are the most useful for cosmological studies, because their properties (e.g., abundance) are the most sensitive to the expansion 
rate and their energetics is dominated by gravity, and less affected by feedback than are less massive objects.

The expected X-ray properties of the PCC are given in Table~\ref{tab:numbers}.  We see a significant number of new clusters, i.e., those
not already contained in the RASS cluster catalogs (REFLEX and NORAS).   Most should have temperatures $T>5-6$~keV and have luminosities
falling in the decade just below the RASS flux limit.  Interestingly, they are easy to see with current X-ray instruments, if one knows were
to look.  And this is our key point: these useful clusters are rare and hence can only be found by an all-sky survey, such as the \pl\ survey.

Our primary conclusion is that extensive X-ray follow-up of the new \pl\ clusters is feasible.  Based on our cluster model, we estimate, for example, 
that \xmm\ could measure the temperature of any newly-discovered \pl\ cluster to 10\% with only a modest 50~ks exposure.
This opens the path to important science with large follow-up programs on \xmm\ and \textit{Chandra}.  A $\sim 5$~Ms program
on \xmm, for instance, could measure profiles and temperatures for $\sim 100$ \pl\ clusters at $z>0.6$, and masses for a large 
subset.  This would be a significant increase in sample size and lead to important advances in the use of clusters as cosmological
probes, as well as in our understanding of cluster physics at higher redshifts.   While ambitious, this type of program is feasible and 
falls naturally under the category of {\em Very Large Programmes} envisaged for \xmm\  after 2010. 

In conclusion, we expect the PCC to be unique for its ability to find rare, massive clusters out to high redshifts, and hence become
a workhorse cluster catalog for many types of detailed cluster studies.  In particular, we expect the PCC to initiate important 
X-ray follow-up programs on \xmm\ and \textit{Chandra}.

\begin{acknowledgements}
We are very appreciative of helpful conversations with M. Arnaud and G. Pratt concerning cluster scaling relations. We are also thankful for the 
very useful comments made by the referee that helped improve the quality of this paper. A. Chamballu and J.G. Bartlett 
also thank the The Johns Hopkins University Department of Physics and Astronomy for their hospitality during part of this work.  A
portion of the research described in this paper was carried out at the Jet Propulsion Laboratory, California Institute of Technology, 
under a contract with the National Aeronautics and Space Administration.
\end{acknowledgements}

\bibliographystyle{aa}
\bibliography{SZ-X-ray_v3}
\end{document}